\documentclass[acmlarge]{acmart}

\usepackage[colorinlistoftodos,prependcaption,textsize=tiny]{todonotes}
\AtBeginDocument{%
  }

\setcopyright{none}

\acmJournal{JOCCH}

\begin{document}

\title{Running a Data Integration Lab in the Context of the EHRI Project: Challenges, Lessons Learnt and Future Directions}

\author{Herminio García-González}
\email{herminio.garciagonzalez@kazernedossin.eu}
\orcid{0000-0001-5590-4857}
\affiliation{%
  \institution{Kazerne Dossin: Memorial, Museum and Research Centre on Holocaust and Human Rights}
  \city{Mechelen}
  \country{Belgium}
}

\author{Mike Bryant}
\affiliation{%
  \institution{NIOD Institute for War, Holocaust and Genocide Studies}
  \city{Amsterdam}
  \country{The Netherlands}}
\email{m.bryant@niod.knaw.nl}
\orcid{0000-0003-0765-7390}

\author{Suzanne Swartz}
\affiliation{%
  \institution{KU Leuven M.Sc. in Digital Humanities, 2024}
  \city{Leuven}
  \country{Belgium}
}
\email{suzannejswartz@gmail.com}
\orcid{0009-0000-8271-6288}

\author{Fabio Rovigo}
\affiliation{%
  \institution{Vienna Wiesenthal Institute for Holocaust Studies}
  \city{Vienna}
  \country{Austria}}
\email{fabio.rovigo@vwi.ac.at}
\orcid{0000-0001-5760-3185}

\author{Veerle Vanden Daelen}
\affiliation{%
  \institution{Kazerne Dossin: Memorial, Museum and Research Centre on Holocaust and Human Rights}
  \city{Mechelen}
  \country{Belgium}
}
\email{veerle.vandendaelen@kazernedossin.eu}
\orcid{0000-0001-7201-061X}

\renewcommand{\shortauthors}{García-González et al.}

\begin{abstract}
Historical study of the Holocaust is commonly hampered by the dispersed and fragmented nature of important archival sources relating to this event. The EHRI project set out to mitigate this problem by building a trans-national network of archives, researchers, and digital practitioners, and one of its main outcomes was the creation of the EHRI Portal, a ``virtual observatory'' that gathers in one centralised platform descriptions of Holocaust-related archival sources from around the world. In order to build the Portal a strong data identification and integration effort was required, culminating in the project's third phase with the creation of the EHRI-3 data integration lab. The focus of the lab was to lower the bar to participation in the EHRI Portal by providing support to institutions in conforming their archival metadata with that required for integration, ultimately opening the process up to smaller institutions (and even so-called ``micro-archives'') without the necessary resources to undertake this process themselves. In this paper we present our experiences from running the data integration lab and discuss some of the challenges (both of a technical and social nature), how we tried to overcome them, and the overall lessons learnt. We envisage this work as an archetype upon which other practitioners seeking to pursue similar data integration activities can build their own efforts. 
\end{abstract}




\maketitle

\section{Introduction}\label{sec:introduction}
Historical research is grounded in the analysis of past events through primary source material, for which archives constitute a cornerstone in facilitating discovery, access, and interpretation \cite{elena2010historical}. One peculiarity of research on the Holocaust in particular is the wide dispersal of sources due to, amongst other things, its vast geographical scope, the intentional destruction of evidence, and the migration of people (and indeed whole populations) before, during and in the aftermath of WWII \cite{speck2014past}. In turn, this dispersal led to the fragmentation of many important primary sources, which are now spread across different countries and custodial institutions, which in many cases use different cataloguing standards and practices. Researchers on the Holocaust at a trans-national level must navigate a patchwork of different archival standards, practices, and technologies \cite{links2016holds}.

The European Holocaust Research Infrastructure (EHRI) project\footnote{\url{https://www.ehri-project.eu/}} was born as a European-funded project seeking to create a network of researchers, archives and digital practitioners in order to actively promote trans-national access, delivering services to the researchers on the Holocaust whether in a digital or physical form \cite{speck2014past}. Starting in 2010, the EHRI project set out to offer a centralised access for archival descriptions held in Europe and beyond. This materialised in 2015 with the launch of the EHRI Portal\footnote{\url{https://portal.ehri-project.eu/}} \cite{blanke2017european} which, up to this date, offers access to more than 380,000 archival descriptions held in 2304 archives across 60 countries.\footnote{Consulted on 21/10/2024} The EHRI Portal follows the International Council on Archives’ (ICA) standards – i.e., the General International Standard Archival Description (ISAD(G)) for archival descriptions, the International Standard for Describing Institutions with Archival Holdings (ISDIAH) for archival institutions and the International Standard Archival Authority Record for Corporate Bodies, Persons and Families (ISAAR (CPF)) for authority records – and integrates a search engine, streamlining researchers’ endeavours and acting as a first access point for trans-national and cross-institutional Holocaust research. Moreover, it also contextualises archival metadata by adding a thematic layer based on the use of a controlled set of terms \cite{alexiev2019semantic} (modelled in the Simple Knowledge Organization System (SKOS) \cite{miles2009skos}) and linking archival descriptions based on shared provenance \cite{erez2020record}.

Consequently, one of the main tasks in the EHRI project is the identification and integration of archival metadata into the Portal. This process has evolved over time. Initially, there was an assumption that it would rely heavily upon the existing use of technical standards for the publication and encoding of metadata – standards such as the Open Archives Protocol for Metadata Harvesting (OAI-PMH) and the Encoded Archival Description (EAD) XML format. In practice, however, this assumption proved optimistic. While there were a handful of cases where EHRI was able to leverage such standards for automated harvesting and ingestion of metadata, this was far from the common case.

Without a viable standards-based approach on offer for the large majority of data providers, other options included manual data entry using the EHRI Portal’s web-based administration interface (developed for cataloguing of material not yet described in electronic form), or the creation of institution-specific bespoke ingestion workflows. Both approaches were recognised as excessively labour intensive and difficult to sustain, the former for the cataloguers, and the latter for technical staff.

Over the course of EHRI’s three phases we have sought to balance the desire to increase the coverage and quality of the metadata in the EHRI Portal with the realities of generally low adoption of or poor support for technical standards, overburdened and underresourced staff, and where the burden lies in the data sharing process – with the data provider, or the integrator. This balance shifted somewhat as the focus expanded in the third phase of the project – EHRI-3 – from the more well-established (and typically well-resourced) data providers in the field, to the long tail of smaller, more variegated, GLAM institutions and even ``micro-archives'', holders of relevant material or collections that do not fit within a particular institutional mould. This ultimately led to the creation of a data integration lab charged with the task of alleviating the big endeavour of setting data integration workflows between data providers and the EHRI Portal. Therefore, in this paper we share our experiences when running the said data integration lab alongside the faced challenges and the lessons learnt from them.

The rest of the paper is structured as follows: Section \ref{sec:relatedWork} reviews the related work, and in Section \ref{sec:ehriPortalAndTools} we introduce the EHRI Portal and its tools for data integration. The data integration lab and the followed methodology are presented in Section \ref{sec:dataIntegrationLab}, while Section \ref{sec:casesAndChallenges} presents some prominent cases and their associated challenges. Section \ref{sec:challengesAndLessonsLearnt} explains the general challenges that we encountered and the lessons learnt from them, and in Section \ref{sec:lookingToTheFuture} we describe how the data integration lab should evolve in future EHRI's phases. Finally, Section \ref{sec:conclusions} draws some conclusions.

\section{Related Work}\label{sec:relatedWork}
Integrating data from different archives for the sake of offering a unified and centralised platform to the user has been tackled by other initiatives including similar – or overlapping – topics. 

National or regional aggregators are becoming the norm in the archival field in an effort to make archives and their holdings more visible and accessible to the end user. They can cover archives without a particular topical focus, like Archiefpunt\footnote{\url{https://archiefpunt.be/}} does for Flanders, or grounded in a specific subject, as for example, Netwerk Oorlogsbronnen\footnote{\url{https://www.oorlogsbronnen.nl/}} covering WWII sources in The Netherlands. One of the main advantages of this approach is the possibility to have a closer collaboration with the partner archives, as well as a better understanding of the local particularities. Moreover, in many cases, there are legal provisions that facilitate this endeavour. On the other hand, their reach is somewhat limited given the geographical – and linguistic – boundaries. They are, however, an invaluable resource in which bigger initiatives can be supported as we explore further in Section \ref{sec:franceArchives}.

Given the complexity of operating in a broader area, trans-national initiatives are far less numerous. Archives Portal Europe\footnote{\url{https://www.archivesportaleurope.net/}} offers a centralised platform to access archival collections all over Europe, though due to its scope there are some inevitable gaps in coverage. In order to overcome this problem – or just trying to accelerate the integration of sources for a specific topic – new aggregators are emerging with a topic-driven mission. In the field of Jewish archival heritage, Yerusha\footnote{\url{https://yerusha.eu/}} offers a portal similar to the EHRI’s one but focused on a broader range of material. As such, it inevitably overlaps with EHRI’s scope, though differs in the acquisition methodology: where in the case of Yerusha it is based on a surveying effort for relevant archival material, EHRI seeks to connect directly with the institutions and integrate their archival descriptions as they are. Ultimately, this makes the EHRI Portal a multilingual platform whereas Yerusha offers its content mainly in English. 

While myriad aggregators may seem like a suboptimal solution for solving the dispersal of sources, they serve to cover small fragments of archives in a specific region or topic which can later be aggregated by supra-aggregators like Europeana\footnote{\url{https://www.europeana.eu/}}, the Common European Data Space for Cultural Heritage\footnote{\url{https://www.dataspace-culturalheritage.eu/en}} or the European Cultural Heritage Cloud\footnote{\url{https://research-and-innovation.ec.europa.eu/research-area/social-sciences-and-humanities/cultural-heritage-and-cultural-and-creative-industries-ccis/cultural-heritage-cloud_en}}. Even though they cover the overarching topic of cultural heritage, and not just specifically archival material, they serve as an entry point for the users who will then be directed to more fine-grained information. In fact, some of these aggregators are already pushing the data to them, like Archives Portal Europe is doing towards Europeana. Furthermore, closer cooperation between the aforementioned aggregators is slowly taking place amidst a general drive for efficiency and to avoid the duplication of effort.

In order to push the data to these platforms some technical adaptations – on account of the different formats used or just to normalise the data – are typically needed. Different aggregators use a variety of techniques to tackle these problems, but in general information is rather scarce prior to committing to the process. As an example, Europeana employs a bespoke format called the Europeana Data Model (EDM) \cite{doerr2010europeana} which is then used as the basis for the representation and ingestion of data\footnote{\url{https://europeana.atlassian.net/wiki/spaces/EF/pages/2059763713/Publishing+guide}}. Archives Portal Europe uses EAD 2002\footnote{\url{https://www.loc.gov/ead/ead.html}} as its base format, with some specific additions\footnote{\url{https://www.archivesportaleurope.net/about-us/join-us/?tab=content-provider}}, and also offers technical assistance in order to convert the content provider’s data to EAD. In both cases, they require institutions to prepare their data in an intermediate format which is then used for ingestion. By contrast, EHRI has adapted its data integration process to one where it – rather than the data provider – takes on the larger responsibility for alignment and conversion of ingested material to the EHRI Portal’s data model.

It is worth noting that some recent initiatives seek to facilitate the reuse of cultural heritage collections by ensuring that they are released in a manner compatible with further computational use. This is the case, for example, of Collections as Data \cite{padilla2019always} which provides a set of guidelines aimed at small- and medium-size institutions on how to publish digital collections (aligned with the FAIR principles \cite{wilkinson2016fair}) and ensuring their further processability by third users. Ultimately, this also allows institutions to retain the control on the data by managing the complete publication workflow which ensures that the CARE principles \cite{carroll2023care} (as a complement of the FAIR ones) are also respected. Nevertheless, while cultural heritage institutions become more technologically independent, leaning on initiatives like Collections as Data for good practices on making their collections computable and by extent reusable, initiatives like the one described here are still relevant.
\section{The EHRI Portal and Its Data Integration Tools}\label{sec:ehriPortalAndTools}
In this section we describe the EHRI Portal, its data model\footnote{\url{https://portal.ehri-project.eu/help/datamodel}} and the existing data integration tools.

\subsection{The EHRI Portal}\label{sec:ehriPortal}
As briefly introduced before, the EHRI Portal gathers archival descriptions relevant to the Holocaust. However, from the very beginning it was deemed necessary to provide more context to the archival descriptions within this overarching trans-national topic. For this purpose, the EHRI Portal uses three main entities (countries, Collection Holding Institutions (CHIs), and archival descriptions). Countries represent an entry point to Holocaust research and provide a historical overview of the country during WWII, its Holocaust history and the general archival situation. From a particular country, the users can search and browse CHIs physically located within its borders. CHIs are described using the ISDIAH standard\footnote{\url{https://www.ica.org/en/isdiah-international-standard-describing-institutions-archival-holdings}}, providing users with contact details and historical and service-related information about the institution. Institutions can hold a varying number of archival descriptions which will be listed in a hierarchical fashion to represent all the possible levels defined by archival practice (e.g., fonds, sub-fonds, series, sub-series, record groups, collections, folders and items) alongside the possibility to host parallel descriptions to accommodate multilinguality. These descriptions follow the ISAD(G) standard\footnote{\url{https://www.ica.org/en/isadg-general-international-standard-archival-description-second-edition}} and, as mentioned before, can contain an arbitrary number of nested descriptions to form a hierarchy. 

While these three entities constitute the core of the Portal, there exists a set of transversal entities that allow for better contextualisation of the archival metadata within the network. Vocabularies (structured using SKOS) define a controlled set of entities for use as archival access points. At the time of writing, there are three such controlled vocabularies in the EHRI Portal: one for subject headings (EHRI terms\footnote{\url{https://portal.ehri-project.eu/vocabularies/ehri_terms}}), and two for places (EHRI ghettos\footnote{\url{https://portal.ehri-project.eu/vocabularies/ehri_ghettos}} and EHRI camps\footnote{\url{https://portal.ehri-project.eu/vocabularies/ehri_camps}}). Similarly, two sets of authorities described using the ISAAR (CPF) standard\footnote{\url{https://www.ica.org/resource/isaar-cpf-international-standard-archival-authority-record-for-corporate-bodies-persons-and-families-2nd-edition/}} define lists of persons and corporate bodies, respectively, which can be linked as the creators of archival material and also as general access points. Finally, all the aforementioned entities can be annotated and linked using a derivative of the OpenAnnotation standard\footnote{\url{https://www.w3.org/community/openannotation/}}. Links can represent temporal, hierarchical, or familial relationships, or those based on provenance, such as establishing that a particular archival description was copied from another collection or institution \cite{erez2020record}.

\begin{figure}[h]
  \centering
  \includegraphics[height=13cm]{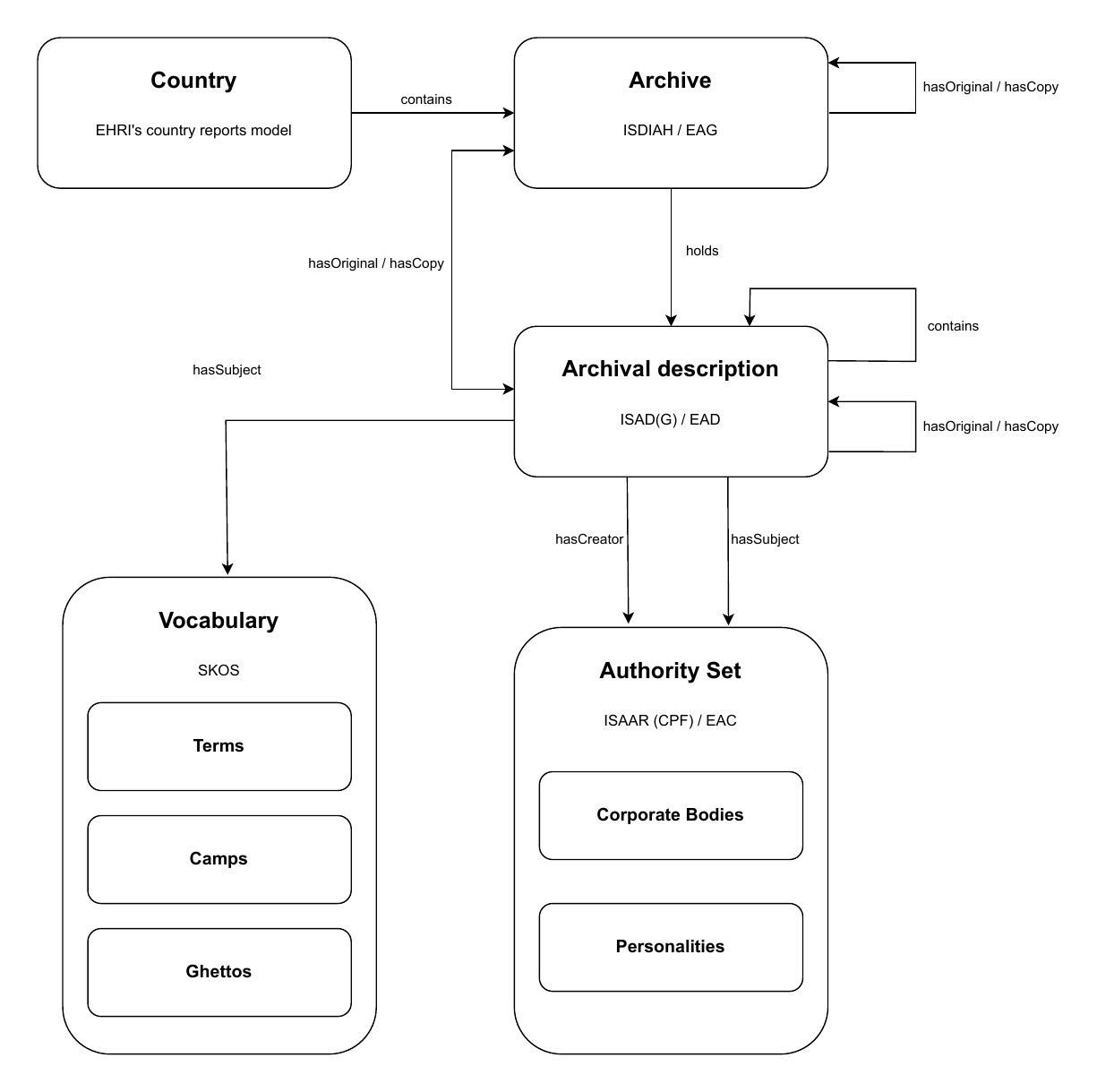}
  \caption{This diagram represents the general data model used by the EHRI Portal where countries use a custom-based model for representing the EHRI country reports, archives use the ISDIAH standard, archival descriptions are based on the ISAD(G), authority sets follow the ISAAR(CPF), and vocabularies are modelled using SKOS. The relations are modelled as follows: countries can contain an arbitrary number of archives, and archives can hold many archival descriptions. Both archives and archival descriptions can be copied or be the original source of other archives and/or archival descriptions. Finally, archival descriptions might be linked to subjects represented in the vocabularies sets or in the authority sets but only entities pertaining to the authority sets can be stated as the creators of an archival description.}
  \label{fig:ehriPortalDataModel}
\end{figure}

A graphical representation of the described data model can be consulted in Fig. \ref{fig:ehriPortalDataModel}.

\subsection{The EHRI Data Integration Platform}\label{sec:dataIntegrationPlatform}
The EHRI data integration tools have undergone a process of evolution and expansion over the course of subsequent project phases, as progressively more was learned about the requirements, the nature of the data held by providers and the ways it was published and/or transmitted, and the approach taken by EHRI in coordinating and managing the ingest of material. Subsequent phases of the project have also enabled us to put more time into making the tools more capable and easier to use, building on what came before.

In EHRI-1, the foundations of the system were put in place with the development of a SAX-based XML processor for importing material into EHRI’s Neo4j-based metadata repository. The basic requirements for this component were that it be:

\begin{itemize}
\item Event-based: to handle large hierarchical XML files in an efficient manner.
\item Transactional: so that validation or other errors did not leave the database in an invalid state.
\item Customisable: so that different profiles of XML (conforming to specific schemas) could be catered to.
\end{itemize}

While the main processor was developed to target hierarchical EAD 2002, the core functionality was generic so that other data types, be it different XML schemas or even tabular data, could be accommodated with the creation of new subclasses. Provider-specific configuration files allowed specific XML paths in the source data to map to different database properties, providing a degree of flexibility to accommodate variations in how individual providers interpreted standards like EAD.

In EHRI-2 this system was further refined, but greater emphasis was placed on conforming the metadata by third parties to a more restricted subset of EAD prior to ingest, using a new GUI-based XML translator. Motivating this change was the desire to help data providers generate valid EAD from what could be a very wide range of primarily XML-based sources, typically exported from some proprietary collections-management system. Named the EAD Creation Tool (ECT)\footnote{\url{https://www.ehri-project.eu/wp-content/uploads/2015/08/D10-1-Collection-Description-Production-Services.pdf}}, it made use of a Google Docs-based tabular mapping system comprised of XQuery expressions. Along with the ECT, a new tool was also developed to help data providers more easily publish their material online using the OAI ResourceSync protocol\footnote{\url{https://www.openarchives.org/rs/1.1/resourcesync }}, named the Metadata Publishing Tool (MPT)\footnote{\url{https://rspub-gui.readthedocs.io/en/latest/}}. The ECT and MPT together allowed institutions \textit{themselves} to both generate and validate EAD, and make it available online, making EHRI’s task much easier when it came to sourcing the material in a sustainable manner and ingesting it into the Portal.

In the third phase of the project, the main objective was to make this system easier to use from end to end. This meant making it more straightforward to track the provenance of material from third-parties, to configure the different types of harvesting methods in use, and to manage the transformations necessary to conform material to the ingest standard. Providing a web-based user interface was also deemed essential, as the collection of command-line tools and API interactions comprising the data integration system in EHRI-1 and EHRI-2 quickly grew overly complex and became difficult to administer.

The first iteration of the EHRI-3 data import tools went live in 2020 and incorporated the three main components of the EHRI ingest pipeline – harvesting, transformation, and import – into the administration backend of the Portal. This system was oriented around datasets defined by a specific Extract, Transform, and Load (ETL) pipeline. Extract, in this case, involves either pulling material from the web (via ResourceSync, Open Archives Initiative
Protocol for Metadata Harvesting (OAI-PMH)\footnote{\url{https://www.openarchives.org/OAI/openarchivesprotocol.html}}, or from a set of URLs), or simply uploading the files manually to the system. Transformation uses the XQuery-based tabular mapping system developed in EHRI-2, allowing transformations to be chained and combined with Extensible Stylesheet Language Transformations (XSLT)-based processes, and incorporates a real-time preview facility allowing mappings to be developed interactively. Finally, the Load functionality encapsulates the EHRI-1 ingest configuration, interfacing with the Neo4j-based database backend.\footnote{The complete documentation for the described data integration suite can be consulted on: \url{https://documentation.ehri-project.eu/en/latest/administration/institution-data/index.html\#}}

This system was later extended to include supplementary aspects of the data ingest process, including the ability to manage how access points (such as subject headings) used by particular institutions were mapped to EHRI’s own controlled vocabularies. Another such process was the ability to run clean-up tasks to handle stale data leftover from past imports, relating to items removed or renamed by data providers. Batch processing functionality was also added, to handle outlier cases where the same ETL pipeline needed to be executed for a large number of datasets (see Section \ref{sec:ushmm}).

Altogether, incorporating the EAD generation tools directly into the EHRI-3 data integration system implied a change of approach relative to EHRI-2, where the strategy was oriented around standalone tools that empowered archives to both generate EAD themselves, and publish it via ResourceSync or OAI-PMH. In reality, this change of approach was more tactical than strategic: while the ECT and MPT tools \textit{did} make the process of generating and hosting conformant metadata easier, in the institutional context they could only do so much, and these tasks still necessitated a relatively high level of technical expertise and IT support to deploy and fully utilise. By integrating the data transformation process into its web-based infrastructure, EHRI could take on some of this burden ourselves (albeit in a manner that was more systematised) and in doing so significantly lower the bar for data providers to participate in supplying their metadata to the EHRI Portal.

\section{The EHRI Data Integration Lab}\label{sec:dataIntegrationLab}
In this section we introduce the EHRI Data Integration Lab, the problems encountered during EHRI’s previous phases that motivated its creation, the methodology followed to set up the different data integration cases and a breakdown of some prominent ones.

\subsection{What Does It Solve?}
Delivering data suitable for integration on the EHRI Portal was a substantial challenge even for the larger, more well-resourced institutions, let alone their smaller counterparts who often lacked the required technical support and expertise. As the EHRI-3 project aimed to cover more thoroughly both the long tail of smaller institutions and so-called ``micro-archives'' (greatly expanding EHRI’s previous reach\footnote{\url{https://cordis.europa.eu/project/id/871111/reporting}}), it was deemed necessary for the project to take on more of the technical aspects involving the integration of archival descriptions, thus making the process more approachable for institutions of all sizes. The mission of the EHRI Data Integration Lab is to manage the creation of data integration workflows for institutions seeking to share their archival metadata, ensuring that this process is as transparent as possible on as technical basis, as well as repeatable and sustainable so that future revisions of the metadata can be smoothly delivered \cite{daelen2015sustainable}.

As discussed in Section \ref{sec:dataIntegrationPlatform}, this shift of emphasis prompted a change on how the data integration tools were operated, making them more manageable by EHRI representatives and supporting a broader range of XML schemas (i.e., harvesting all kinds of XML files instead of just EAD). While this did not prevent institutions from setting up their own workflow using the new environment, they were not forced to do so and could delegate this to the data integration lab. At the same time, the tools offered increased productivity, as staff could be trained on more specific activities in a common environment, and resources better shared from one case to the next. The goal was to avoid as much as possible EHRI-specific \textit{ad-hoc} development within institutions’ own IT infrastructures, but instead using the mechanisms already put in place by the institutions for data exchange (API, exports, etc.), resulting in a more future-proof solution.

Initially, the data integration lab was conceived as a mobile lab, offering the possibility to visit the institutions and deliver the technical set-up \textit{in situ}. While this possibility was offered to institutions, its reach was somewhat limited due to different circumstances like the breakout of the COVID-19 pandemic or the travel restrictions on the following years. Nevertheless, in some cases, it was possible to exercise this option, but from our experience and the achieved results, we cannot correlate a visit to the archive with a better outcome on setting up a data integration workflow. It is possible that this can be motivated by the fact that people got more used to work and collaborate in virtual environments after the pandemic for which we do not gather enough evidence from this project to totally assert it.

\subsection{Methodology}
In order to make this work more structured and manageable for the data integration team members, a methodology was put in place. This methodology defines a central access point for filing data integration requests to the EHRI project in the form of an online questionnaire\footnote{\url{https://forms.gle/YFCEhJzkEkSVT7yt9}}, ensuring that all relevant key aspects about the provider’s technical infrastructure are collected beforehand. Moreover, internally, it keeps the information about all the cases centralised and traceable. These requests are then integrated into a management system which provides traceability of the individual tasks and allows for an efficient organisation within the team. Data integration workflows are firstly implemented under a staging environment so that a representative of the institution can verify if the data has been integrated correctly and ask – if required – for changes to the data. Only after receiving the final approval by the institution the data is moved to the production portal and consequently made publicly available. If at any point in time, an update is required, the described process will be replicated, ensuring that institutions are always in control of how their data is represented on the EHRI Portal.

In an effort to provide a transparent data integration workflow that aligns with the FAIR principles, the resources used for integrating the data of a specific institution are shared in a public repository\footnote{\url{https://github.com/EHRI/DataIntegrationLabResources}}. This repository contains a folder for each of the implemented cases and within each folder one can find the resources (i.e., mapping rules, configuration files and links to the datasets) employed for this case alongside a README with instructions on how to replicate the full workflow.

\subsection{Legal Background}
Archival finding aids are creative works resulting from intellectual effort by an archivist or group of archivists, who will study the material and describe it according to some general rules. As such are subject to the same intellectual property rights (IPR) as other types of material, and this means that any re-utilisation of the descriptions - modified or not - should respect the IPR of the authors. It is worth noting that a lack of licence on publicly available content does not mean that it can be used for free or openly but rather it is normally understood to be protected by an exclusive copyright, even though this consideration can vary greatly depending on different countries’ laws.

Archival descriptions pertaining to the Holocaust can also involve, in some rare cases, the processing of special categories of personal data according to the GDPR (Art. 9) which need to be taken into account when including them in public archival descriptions (as safeguarded by Recital 158). As a European project, EHRI's operations need to be aligned with the GDPR, and any possible breach by the project or one of its partners should be communicated expeditiously in order for it to be resolved.

To provide a legal framework for the work of the data integration lab that could cover, amongst other things, situations relating to IPR and data privacy, EHRI developed a text named the Content Provider Agreement (CPA). This agreement covers the exchange of data between the provider and EHRI, as well as the representation of the descriptions, their hosting, and their possible reuse during all the phases of the project. It also comes with a set of compromises for both parties regarding the fulfilment of the GDPR and intellectual property rights.

While these sorts of agreements are quite commonplace in business to facilitate operations involving IPR, during the EHRI-3 project it has proved to be more challenging than anticipated, even more so, in some cases, than the technical challenges described in this paper. While signing the CPA was a straightforward formality for some institutions, others had minor questions or needed additional clarifications. At the other end of the scale were those institutions who felt they needed to undertake legal consultation, which in the most challenging cases could end up requiring an amendment of the CPA. In some cases, the CPA could not be signed, which prevented the relationship between EHRI and the data provider moving forward. In Section \ref{sec:nonTechnicalChallenges}, we further explore some of the actions put in place in order to mitigate the issues derived from this.
\section{Some Highlighted Cases and Their Companion Challenges}\label{sec:casesAndChallenges}
In this section we provide an overview of some cases that will help illustrate the reach and scope of this data integration lab, as well as some of the challenges encountered.

\subsection{United States Holocaust Memorial Museum (USHMM)}\label{sec:ushmm}
USHMM\footnote{\url{https://www.ushmm.org/}} is one of the biggest international aggregators and collectors of Holocaust-related material, with substantial collections in its archives, museum, and audio-visual library. In sharing metadata about its holdings with EHRI, challenges were encountered due to the size of this dataset, consisting of many thousands of hierarchical collections and item-level descriptions, and different approaches were adopted as the data was updated and refined in successive project phases. 

For EHRI-3, the new ingest environment was used and the USHMM export provided an ideal test-case for a large and complex dataset, necessitating a number of refinements and new accommodations. Firstly, the full dataset had to be split into series by accession year in order to make the ingest process more manageable for the EHRI Portal's backend by reducing the size of single transactions. This, in turn, necessitated the creation of a batch processing system to automate the execution of the same ETL process over a large number of datasets. 

It is worth noting that the USHMM was the only institution that preserved its EHRI-2 workflow involving the publication of data using the ResourceSync protocol, but dispensed with the conversion to EAD, instead publishing its bespoke XML data directly with the conversion taking place on EHRI’s servers. This ultimately allowed them to adjust more straightforwardly the fields that were exported, leading to richer descriptions on the EHRI Portal and helping ensure more timely updates\footnote{The results can be found on: \url{https://portal.ehri-project.eu/institutions/us-005578}}.

\subsection{Center for Urban History}
The Center for Urban History of East Central Europe\footnote{\url{https://www.lvivcenter.org/en/}}, based in Ukraine, possesses a rich digital archive related to the Holocaust containing interviews, maps and photos. For various reasons pertaining to the software capabilities, the export of the items was only possible in a CSV format. As explained in Section \ref{sec:standardsCoverage}, this was a situation common to many institutions around the world. 

Prior to this case, a workflow was put in place to convert Excel files to EAD in order to integrate data on the EHRI Portal from other institutions.\footnote{\url{https://blog.ehri-project.eu/2022/04/25/converting-from-excel-to-ead-xml/}} This workflow was adapted to this specific case, taking into account one particular aspect, the existence of two hierarchical levels within the CSV file that had to be incorporated in different hierarchical levels\footnote{The results can be found on: \url{https://portal.ehri-project.eu/institutions/ua-006557}}. Fortunately, each row incorporated information from the broader collections making the conversion to EAD easier than it usually is in these kinds of cases.

Nevertheless, the process faced many setbacks stemming from the inconsistency between English and Ukrainian descriptions, which made the process very error-prone and time consuming. Throughout the process the provider needed to update their data to correct mistakes that were uncovered, and invariably some trial-and-error was involved. This case made us realise that, while sometimes this method can be the only way forward to import data, alternative methods like that developed for the Vilna Gaon Museum of Jewish History (see Section \ref{sec:vilnaGaon}) may provide a faster and more sustainable way to integrate tabular datasets.

\subsection{Ottawa Jewish Archives}
This institution\footnote{\url{https://www.jewishottawa.com/ottawa-jewish-archives}} holds material related to the Jewish Community life in Ottawa, of which a selection is relevant to the Holocaust. While the Ottawa Jewish Archives staff were able to provide EHRI with an export of the Holocaust-related material in XML format, the hierarchies were broken up such that the fonds level description and the item level units it contained were represented by separate XML files. Therefore, it was necessary to reconstruct the hierarchy in order to present these descriptions in a coherent manner.

From a practical point of view this was implemented on the EHRI Portal\footnote{\url{https://portal.ehri-project.eu/institutions/ca-006572}} using two distinct datasets: one for the fonds level and another for the item level. They were then executed in a specific order with the item level descriptions ingested first, embedded in an EAD containing the hierarchy as a skeleton, consisting of the fonds level identifier and title, as extracted from the items’ attributes. Following this, the full fonds metadata was ingested, enriching the skeleton descriptions created in the item level import.

This workflow enables the relationship between different levels in order to reconstruct a hierarchy within the EHRI Portal when the data is coming from different files, assuming a identifier reference to the parent is included in lower level descriptions. It is a generally more sustainable approach than doing the same procedure via bespoke scripting and it is flexible enough to cover full hierarchies. The main drawback in practice was the amount of time needed to execute the ETL for different levels.

\subsection{Vienna Wiesenthal Institute for Holocaust Studies}
The Simon Wiesenthal Archive (SWA), which comprises approximately 200 linear meters of materials, is the largest collection within the Archive of the Vienna Wiesenthal Institute for Holocaust Studies (VWI)\footnote{\url{https://www.vwi.ac.at/index.php/en/documentation/archive/simon-wiesenthal-archive}}. Integrating the SWA's metadata into the EHRI Portal posed a technical challenge, as a complete export in a single file, particularly one that preserved the hierarchical structure, could not be achieved. Only a portion of the SWA's archival descriptions could be exported as EAD files that contained the hierarchical structure. Others were available only in a custom XML format, which, although it typically contained more detailed information, did not include the hierarchy. The solution was to import metadata in both formats, prioritising the hierarchy-preserving EAD-XML files where available, and relying on the custom XML format when it offered additional information not present in the EAD version.

This amalgam of files was imported into the respective series using the tools available on the EHRI data integration platform. Two collections were imported down to the item level in a single step, using the hierarchical export in EAD format. For the remaining collections, the lack of hierarchical information in the custom XML format resulted in the need to import each level separately. Therefore, the utilisation of two distinct import formats enabled the incorporation of the metadata of the entire SWA\footnote{The complete results can be consulted on: \url{https://portal.ehri-project.eu/units/at-006006-vwi_swa}}, albeit with variable levels of detail, and represents precisely the kinds of compromises that have to be reached in order to effectively import the data of an archive using the existing technologies.

\subsection{Wiener Library Tel Aviv}\label{sec:wienerLibrary}
The Wiener Library Tel Aviv\footnote{\url{https://en-cenlib.tau.ac.il/Wiener}} is a special section of the Tel Aviv University's Library consisting of a library and archive about the Holocaust. Due to this very specific arrangement within a larger library, the software in use is one specifically dedicated to libraries, i.e.,  Ex Libris Alma. While using a library-specific set-up for an archive can lead to some misalignments resulting from differing terminology and standards, it was not the biggest challenge in this case due to the flexibility of the overall system.

At first, the representatives were able to export EAD files from the different collections but the nested levels of the hierarchy contained less details than those present in the system. This was probably due to the fact that the system is not centred around archival standards, despite some support for them. However, due to the great flexibility and the variety of supported data exchange formats\footnote{\url{https://knowledge.exlibrisgroup.com/Alma/Product_Materials/050Alma_FAQs/General/Standards}}, we were able to discover an open endpoint supporting Linked Data. Based on this, we were able to automatically harvest a Bibframe RDF/XML\footnote{\url{https://developers.exlibrisgroup.com/alma/integrations/linked_data/bibframe/}} representation of the lower levels, containing much more detailed information about those records. Moreover, this endpoint allowed us to establish a sustainable connection with the provider, removing the need to manually export and transmit the data\footnote{The results can be found on: \url{https://portal.ehri-project.eu/institutions/il-002820}}.

\subsection{Yad Vashem (YV)}\label{sec:yadVashem}
YV\footnote{\url{https://www.yadvashem.org/index.html}} is another of the largest collectors and aggregators of Holocaust-related material in the world. This made the integration of their metadata of great importance for the EHRI project since its inception. Nevertheless, in previous EHRI phases, the import of their collections was based on a custom-made Access database (exported from their main system and thoroughly curated by YV archivists for EHRI) containing the full archival hierarchy which was then converted to EAD using an \textit{ad-hoc} script\footnote{\url{https://blog.ehri-project.eu/2019/07/03/integrating-new-data-methods-practice/}}. 

Unfortunately, when the EHRI-3 data integration lab had to update these descriptions the situation was rather different. This labour-intensive process of curating a bespoke export could no longer be supported, as it was based on the previous project financing no longer in place. Due to the very specific nature of the export and its disconnection from the main database – all the more so after several intervening years – it was considered prohibitively difficult to recreate the full process. Moreover, on EHRI’s side the script was no longer available, nor was the staff expertise. As a more permanent solution, the data integration lab and YV’s IT department agreed to create a bespoke API that could export the collection level to EHRI. This led to the update of the collection level\footnote{\url{https://portal.ehri-project.eu/institutions/il-002798}} but unfortunately, the recreation of the full hierarchy, and therefore its full update, was not possible to achieve due to time-constraint reasons on YV’s and EHRI’s side and is set to be resolved in future phases.

\subsection{Vilna Gaon Museum of Jewish History}\label{sec:vilnaGaon}
This institution\footnote{\url{https://www.jmuseum.lt/en/}} is a museum that, like many others, also holds an archive within. In this case, many different approaches were explored but none of them led to a workable and sustainable solution. However, as a Lithuanian institution, they are legally obliged to deliver their data to LIMIS\footnote{\url{https://www.limis.lt/}}, a national aggregator for museums’ collections. As a national aggregator, LIMIS has a much more open and interoperable platform, supporting a range of different formats including an API delivering JSON or XML. Our first try was to make use of the XML outputs given the EHRI Portal’s XML-centric platform but this format proved to be less rich in detail than its JSON counterpart.

As mentioned earlier, the EHRI Portal’s data integration tools do not, at the time of writing, support JSON inputs and conversion. However, based on a parallel process to deliver the Portal’s data as a Knowledge Graph (KG) \cite{Garcia-Gonzalez23}, and previous experiences converting RDF/XML to EAD (see Section \ref{sec:wienerLibrary}), we decided to include a pre-transformation process able to convert JSON to RDF/XML using declarative mapping rules \cite{van2023declarative}. Declarative mapping rules allow converting heterogeneous data sources to an integrated RDF file in a more flexible and reusable way than most \textit{ad-hoc} methods, and in practice we only employed a minimal set of rules in ShExML \cite{Garcia-Gonzalez20a} (a language developed by one of the authors of this paper) to convert the JSON input into RDF/XML. In order to integrate this into the existing workflow, a web service was created encapsulating the invocation of the ShExML engine and making it callable from within the URL-based harvester.

\begin{figure}[h]
  \centering
  \includegraphics[width=\linewidth]{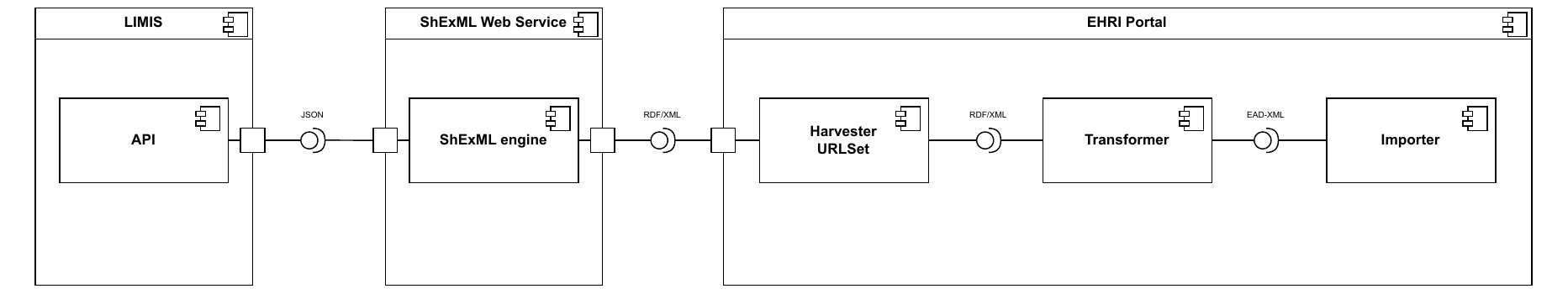}
  \caption{Diagram of the system implemented to transform JSON files harvested from an API into EAD using the ShExML engine, prior to the standard EHRI Portal data transformation workflow.}
  \label{fig:workflowVilnaGaon}
\end{figure}

After this pre-transformation step, the EHRI Portal receives a set of RDF/XML files that can be converted to EAD using the standard tools and ingested without further additions\footnote{The results can be found on: \url{https://portal.ehri-project.eu/institutions/lt-002881}}. A graphical representation of this workflow and its components can be seen in Fig. \ref{fig:workflowVilnaGaon}. This experimental set-up not only solved the integration of data for this institution but also served as a use case to demonstrate how EHRI’s data integration tools can evolve in the future to cope with more heterogeneous data formats.

\subsection{FranceArchives, a National Aggregator Experience}\label{sec:franceArchives}
The FranceArchives\footnote{\url{https://francearchives.gouv.fr/}} aggregator includes metadata for collections and fonds from archives throughout France. The site went live in 2017, and since then has amassed over 26 million archival descriptions\footnote{Consulted on 06/03/2025}. These records span fifteen centuries and a wide range of topics, and come from both national and regional-level archives. Collections that may not have previously been easily accessible, particularly at smaller or regional archives, are also included, making it a valuable resource for researchers. The site also provides different methods for searching, querying and exporting the metadata, including a general keyword search, a SPARQL endpoint, or browsing by theme. 

One of the key aspects of the integration task was the identification of Holocaust-relevant collections – in contrast to those cases where the CHIs only hold Holocaust-relevant descriptions – for which the use of the site’s search features was therefore essential \cite{SwartzMScThesis}. As a result of intensive keyword and theme-focused searches on the site, approximately 280 collections potentially relevant to the EHRI Portal were identified.  Nevertheless, as an aggregator, the FranceArchives portal accepts metadata in multiple formats and standards from institutions interested in sharing their collection metadata. Some metadata may come through in hierarchical formats such as XML, whereas other archives may provide finding aid PDFs. As such, the level of detail of the metadata can vary significantly. This lack of standardisation can make integration challenging as to whether incorporating the entire collection in the EHRI Portal, or filtering within a larger collection if only a selection of subfolders is relevant to the EHRI Portal.

The platform initially began with EAD as its main standard, and released an OAI-PMH endpoint for metadata harvesting purposes. In 2022, however, it shifted to the Records in Contexts Ontology (RiC-O)\footnote{FranceArchives, ``Rejoindre le portail'', \url{https://francearchives.gouv.fr/fr/about}, accessed 20 January 2025, and FranceArchives, ``Records in Contexts: un nouveau modèle de description archivistique'' \url{https://francearchives.gouv.fr/fr/article/334841641}, accessed 20 January 2025} making the OAI-PMH endpoint potentially deprecable in favour of data exchange standards more amenable to the Semantic Web \cite{Berners-Lee2001semantic}, like the existing SPARQL endpoint.  Nevertheless, due to the current EHRI Portal data integration platform, the OAI-PMH endpoint (and its companion EAD format) was more compatible overall, requiring small XSLT transformations rather than fully transforming RiC-O to EAD.

Incorporating collection metadata from FranceArchives was initially devised with the goal of developing strategies and methods to integrate and update collections from similar aggregators in the future, helping us cover a national landscape in a much more efficient manner due to leaning on previous efforts and leveraging a unified technical stack. The FranceArchives process differs from that of archives EHRI has worked with directly as it does not have the same level of continuous human communication and troubleshooting that communication with an individual archive would, helping to determine which collections to share with EHRI, how to share that metadata using guidance from the EHRI data integration team, as well as jointly identifying any challenges or integration glitches. 

While a benefit of FranceArchives is that its open licence\footnote{FranceArchives, ``Open data - Données ouvertes'' \url{https://francearchives.gouv.fr/fr/open_data}, accessed 20 January 2025.} removes potential non-technical, bureaucratic challenges of metadata sharing (metadata can be shared as long as attribution is given to FranceArchives), fine-tuning the data identification becomes a challenge because of the aforementioned missing human connection to the archive. Additionally, the FranceArchives portal continues to evolve, and therefore a continuous exploration of sustainable methods for locating and updating EHRI Portal-relevant collections becomes of the utmost importance. 
\section{Challenges and Lessons Learnt}\label{sec:challengesAndLessonsLearnt}
In this section we describe the encountered challenges divided by categories and some of the most notable lessons learnt from them. 

\subsection{Standards Coverage, Limitations on Data Exchange and Data Governance}\label{sec:standardsCoverage}
While the Cultural Heritage field, and more specifically that of archives, have produced a set of standards covering different parts of the data exchange workflow (e.g., OAI-PMH for harvesting metadata or EAD for encoding archival finding aids), the support for them – as noted in Section \ref{sec:introduction} – is rather inconsistent across institutions and the systems they use. In some cases, organisations have a strong policy of providing wide support for different standards (as for example the case of Ex Libris Alma described in Section \ref{sec:wienerLibrary}) whereas in other cases, if the data is accessible at all it is via a bespoke format or API, which inevitably makes the data integration process more complex. This, ultimately, makes the exploratory phase of data integration a process in which different possibilities have to be weighed, taking into account how they will affect the final results, as well as their potential repercussions for the future-proofness of the solution that is delivered.

IT service providers are a related factor, particularly with regard to the data exchange policies they wish to implement and support. Given that they may have other use-cases or product roadmaps in mind, the implementation of standards and other robust data exchange methods does not always have the highest priority alongside other business functionality, and experience shows us that many institutions are unable to export their data except in extremely limited ways as a result. Although it may seem surprising, staff working with collections management systems often do not seriously consider questions of interoperability until they want to collaborate in a project such as EHRI, long after IT procurement decisions have been made.

This naturally raises concerns about data governance, and whether institutions have the right to reproduce their own data as needed, or even migrate from one collections-management system to another. During our interviews with different institutions we have detected a tension between the priorities of private IT service providers seeking to maximise the adoption of their solutions in the Cultural Heritage sector, and the mission-driven activities of GLAM institutions who want to be able to manage their data as freely as possible. While in many cases the functionality requested by institutional clients is covered by the solutions delivered by their IT service providers, it is at the moment of a migration to another system or when a collaboration with an aggregator, like the EHRI Portal, wants to be pursued, that these tensions come to the fore. While it is not the purpose of this paper to give solutions to this problem, we would like to highlight two general recommendations that could be applicable to GLAM institutions and service providers alike: 1) ensure that the data introduced in one platform can be exported in reusable formats out of the platform following industry standards (e.g., archival descriptions in EAD) and 2) state these requirements in the Service Level Agreement (SLA) or similar binding contract, especially in cases where the work is funded by public money and should therefore be available for re-use by third parties.

\subsection{Hierarchies Not Always Considered First-Class Citizens}\label{sec:hierarchiesProblems}
Even in the cases where multiple data exchange formats and publication possibilities are theoretically supported, additional challenges still appear. A very prominent one that we have encountered across many institutions is inadequate support for describing archival hierarchies. While the most prominent archival conceptual standard – ISAD(G) – describes how descriptions can be hierarchically organised, the implementation of this fundamental aspect across different systems reflects a totally different reality and in many cases the flat tabular form derived from the relational model, in which many applications’ databases operate, is imposed on the application. This is not only true for the user interface but it is also translated to the export functionalities where, even when providing hierarchical data formats like JSON and XML, the systems do not represent the full hierarchy but rather a direct translation of the tabular form. One of the main drawbacks of using a tabular model to represent hierarchical data is its inability to easily capture the hierarchical relationships, leading to complex JOIN clauses, hierarchy reconstruction via \textit{ad-hoc} scripting, or storing data in ways that violate database normal forms. Moreover, in some cases, the hierarchical information is simply not included in the export.

\subsection{Heterogeneity of Formats as a New Reality}
A related problem for EHRI has proved to be the heterogeneity of formats in which the files can arrive for ingest into the Portal. While the EHRI Portal data integration tools were initially based around the idea of XML as the \textit{de facto} standard in the Cultural Heritage field, also motivated by the use of EAD as the principal transmission format for archival descriptions, recent technological trends \cite{breje2018comparative} have made this pragmatic choice harder to sustain.

As mentioned in the previous section, some cases required the adaptation of the existing workflow to handle non-XML formats with a varying degree of usability and sustainability. For example, the conversion of Excel files to EAD helps to solve an immediate problem and brings more collections to the EHRI Portal. However, when applied to larger collections, and those not homogeneously shaped, the process was typically quite inefficient, error-prone, and time consuming where, in the worst cases, a mapping error at one stage necessitated rerunning the whole workflow.

Conversely, the inclusion of a pre-transformation step using declarative mapping rules like ShExML (as in the case of the Vilna Gaon Museum of Jewish History), proved to be much more flexible and sustainable, and could be more easily integrated into the existing ETL pipeline. Altogether, it is still an experimental use-case, and as such adds inefficiencies by having to convert the input twice, first to RDF/XML and then to EAD. Integrating these technologies more deeply, however, would require a more substantial change to the EHRI data integration environment, mainly accepting more formats for ingest other than just EAD. Such changes are increasingly pressing, however, with the advent of the first stable release of the ICA’s Records in Contexts Conceptual Model (RiC-CM)\footnote{\url{https://www.ica.org/resource/records-in-contexts-conceptual-model/}} and RiC-O\footnote{\url{https://www.ica.org/standards/RiC/RiC-O_1-0-2.html}}, which supersede ISAD(G), ISAAH-CPF, and ISDIAH (and their EAD, EAC and EAG counterparts) in providing a model for archival metadata.

\subsection{Exploring a Federated Approach for Aggregators}
Broadly speaking, the work of sustaining an aggregation platform like the EHRI Portal is never going to be finished, as new institutions and their materials continuously need to be identified, integrated, and later on updated to ensure the future relevance of the hosted data. In this respect, the process can seem like a Sisyphean task in trying to achieve a sustainable level of operation. Even when relying on existing national aggregators like the cases developed with LIMIS and FranceArchives, there still exists a substantial operational cost in keeping all this metadata up-to-date and ensuring its accuracy.

While this picture might seem somewhat negative, it is a challenge that prompts discussion for alternative approaches and for exploring new methods of interoperability in the Cultural Heritage field. In this regard, in one of our previous works \cite{Garcia-Gonzalez23} we have explored the possibilities that the Semantic Web technologies – and their advocacy for shared vocabularies, unambiguous identifiers, and graph-shaped data over a decentralised web – can bring to this problem, specifically how creating a KG of the EHRI Portal’s data can alleviate data integration endeavours by making more lightweight integrations of metadata that can be retrieved on-the-fly from data providers serving their data in compatible standards. In addition to removing several current barriers to effective data integration, it could also lessen the burden data providers currently face in (often manually) pushing their data to different aggregators. 

At the same time, this would have a disruptive impact on how aggregators operate nowadays: some of the effort currently put towards centralisation of data could be diverted into maintaining the connections across the federated network and further contextualising the collections amongst them, like EHRI has been doing by means of the EHRI vocabularies (for enabling thematic searches) and provenance-based linking (providing more contextual information to researchers). This is not new to Cultural Heritage where standards like IIIF\footnote{\url{https://iiif.io/}} \cite{snydman2015international} have taken advantage of this federated approach to make images interoperable across platforms or, in recent years, initiatives like the European Cultural Heritage Cloud\footnote{\url{https://research-and-innovation.ec.europa.eu/research-area/social-sciences-and-humanities/cultural-heritage-and-cultural-and-creative-industries-ccis/cultural-heritage-cloud_en}} and the Common European Data Space for Cultural Heritage\footnote{\url{https://www.dataspace-culturalheritage.eu/en}} have emerged to solve part of these problems. Nevertheless, due to the recent appearance of the first stable version of Records in Contexts (RiC), the required technical underpinnings are now available in the archival field and hence this realisation seems more achievable than before.

\subsection{Non-technical Challenges}\label{sec:nonTechnicalChallenges}
Even though most of the issues raised in this section pertain to the technical aspect, there are some challenges related to organisational issues or the actual historical content.

Whereas there are many GLAM institutions which are dedicated to memorialising and facilitating research into the Holocaust, these are only a minority of those which have custody of Holocaust-related material. This creates an organisational challenge by which it is necessary to determine what is within EHRI’s scope prior to importing such institution’s metadata into the EHRI Portal. Typically, the bigger the archive, the more complicated this task is. As an example, state or national archives contain multiple collections dedicated to various different topics of interest for a specific nation. Of those, only a small portion will be dedicated to the Holocaust, and in the worst cases, only some files or individual documents will be relevant. In the latter case, this complicates even more the required filtering, and later ingest, into the EHRI Portal. This mainly constitutes a content challenge in which an archivist, with a great knowledge about a specific collection, should decide what is relevant, and which archival level or levels would be more interesting to be represented on the EHRI Portal, delivering fine-grained information but sufficient contextual details. Unfortunately, this is not always possible as this specialised archivist may no longer work for the organisation – or simply by a lack of time, let alone the cases in which a specific collection has not yet been described.

Another challenge involves communication with the prospective archives. On many occasions, they understand the benefits that the EHRI Portal can provide in terms of visibility, but find the technical approach too challenging, or are uncertain about how their collection metadata might end up being displayed in the EHRI Portal. In response, we have tried to highlight that the data integration lab would take care of the technical aspects and that all the data would need their approval before being publicly available (via a test integration in the staging environment). However, sometimes this was not sufficient and the archives were interested in more short-term benefits from the participation in EHRI, beyond the enhanced visibility. In that regard, we tried to offer immediate consultancy on technological aspects with a dual purpose: helping them to solve or analyse their technical problems, and making the data exchange with the EHRI Portal possible or more feasible. Amongst other topics, we covered: Search Engine Optimisation (SEO), data normalisation, controlled vocabularies, open source solutions (like Access to Memory\footnote{\url{https://www.accesstomemory.org/}}) and Linked Open Data (LOD). In addition, more archival-related expertise was also shared, closely related to the EHRI Experts Lab \cite{dilman2025ehri}, covering topics such as: digitisation, names records and their indexing, geographical representation, etc.

However, in terms of communication, one of the most common roadblocks was related to the signing of the CPA. As noted above, the CPA was a necessary prerequisite in order to have a reliable legal framework for data exchange and the subsequent hosting of metadata on the EHRI Portal. This was one of the biggest challenges that the data integration lab had to deal with as this legal text often raised many questions and concerns, also fuelled by the trans-national aspect of the project involving archives in different countries with very different laws relating to this area. In order to mitigate these problems, and in collaboration with others in the project, we decided to introduce the CPA into the conversation from the very beginning, and at the same time offered a separate session with legal experts in which the CPA was more thoroughly introduced, alongside its necessity and the clarification of some of the more controversial articles. Nevertheless, in some cases, the whole signing process could take up to one year, involving many different meetings and several amendments. This demonstrated a clear need for a more careful communication about the legal aspects, similar to that done with the technical counterpart and described in this paper.

It is worth highlighting that these issues are outside the technical scope of this data integration lab but, nevertheless, they greatly impacted the outcomes it was able to achieve. Therefore, it should be noted that transversal aspects, like the legal one, but also possible social, cultural, and institutional factors, should be addressed and considered in the methodology from the outset, as they have been shown to have an enormous impact on the overall effectiveness of the technical endeavour.

\section{Looking to the Future}\label{sec:lookingToTheFuture}
This paper has gathered our experiences running a data integration lab during the EHRI-3 project to enrich the metadata present on the EHRI Portal about Holocaust-related archival material. However, the end of this project does not mean that EHRI's data integration work has concluded. On the contrary, as mentioned before, this task is by nature an on-going one. Nevertheless, after three phases of running as a European-funded project, the EHRI project has recently transition to a permanent organisation\footnote{\url{https://www.ehri-project.eu/inauguration-ceremony-in-warsaw-ehri-becomes-an-eric-to-secure-the-future-of-holocaust-research/}} via its establishment as a European Research Infrastructure Consortium (ERIC)\footnote{\url{https://research-and-innovation.ec.europa.eu/strategy/strategy-2020-2024/our-digital-future/european-research-infrastructures/eric_en}}, funded instead by participating countries. While it is not the objective of this paper to detail how an ERIC functions or how the particular EHRI-ERIC will operate, we will try to sketch what this entails for its data integration activities.

Generally speaking an ERIC is constituted by different countries (commonly known as national nodes) and a coordinating body (commonly known as the central hub). The central hub should normally ensure the coordination of the key activities as well as providing some essential infrastructure, though this is not predetermined and can vary greatly between different ERICs. In the specific case of EHRI-ERIC, we envision the creation of federated data integration labs that could ensure a better understanding of national archival landscapes, together with easier follow-up and collaboration with them. It is worth noting that while these days English tends to be used as the \textit{de facto lingua franca}, and to the extent of our operations it worked fairly well, there are still many archives to which EHRI was unable to establish effective communication with due to idiomatic barriers. The national nodes have an opportunity to improve this situation with their geographical and linguistic advantages, increasing the scope and coverage of the EHRI Portal amongst hitherto neglected GLAM institutions. 

Nevertheless, another challenge arises from this new set-up on how to maintain the trans-national aspect of data integration, notwithstanding that the national nodes are going to have a more local approach, and not all the countries will be represented in this next phase. In this sense, the consortium will have to devise a federated approach in which the national nodes would have the autonomy to add new records to the Portal, reproducing their national Holocaust archival landscape, but also taking shared responsibility for the addition of archival records outside their own jurisdiction. In this sense, this federated approach opens up new possibilities of further integrating archival descriptions from many archives around the world but it also presents a set of organisational challenges to preserve the homogeneity of the EHRI Portal’s data, to ensure an even curation and data quality across it, and to oversee that the data integration workflows are developed to the same technical standards and their long-term sustainability is ensured. Inevitably, the central hub will have to formulate guidelines on how these operations should work and seek to coordinate the addition of this new metadata. 

In this sense, we see the work developed by the data integration lab during the EHRI-3 project, and described in this paper, as a blueprint on how future federated data integration labs could work; further underlining what the possible lines of improvement are upon our challenges and lessons learnt.
\section{Conclusions}\label{sec:conclusions}
In this paper we have introduced the topic of data integration of Holocaust-relevant archival metadata into the EHRI Portal as part of the EHRI-3 project. The EHRI data integration lab was established in order to undertake the technical aspects of incorporating metadata from institutions large and small, and with great variation in their ability to provide structured and standardised archival descriptions. Because in EHRI-3 the emphasis in data integration shifted from the most prominent holders of Holocaust-related material to the long tail of varied and often less resourced GLAM institutions and micro-archives, the technical strategy also shifted, adopting a more flexible \textit{take it as it comes} approach in response to the relative dearth of support for standardised formats in the wild. This shift, however, built upon the tools developed in earlier phases of the project, albeit within a more systematised framework and abetted by a new web-based user interface and ETL management environment.

We have described the methodology of the data integration lab, and how the process of working with individual data providers progressed from initial contact, to iterative testing of ETL via the staging environment, to finally metadata being incorporated into the production environment. We have also described the CPA and its purpose as the legal framework underpinning the transfer of metadata and its publication in the context of relevant EU and international laws such as the GDPR.

The various cases highlighted above describe some of the challenges encountered in our data integration efforts, further elaborated in Section \ref{sec:challengesAndLessonsLearnt}. Amongst these issues were difficulties arising from scale, from interpretation of standards or their lack of coverage, from patchy support for hierarchical archival description on the part of data providers (or their cataloguing systems), and the heterogeneity of metadata formats in use, relative to the preponderance of XML across the standards. 

Finally, we offer a set of watchpoints for the most frequently occurring issues, as well as some guidelines on how they can be solved or mitigated in the form of lessons learnt. As such this paper can serve as a blueprint on how similar initiatives can base their data integration activities while also setting the foundations for further data integration endeavours in the context of the newly-funded – and inherently distributed – EHRI-ERIC.

\begin{acks}
This work has been carried out in the context of: the EHRI-3 project funded by the European Commission under the call H2020-INFRAIA-2018-2020, with grant agreement ID 871111 and DOI 10.3030/871111; and the EHRI-IP project funded by the European Commission under the call HORIZON-INFRA-2023-DEV-01, with grant agreement ID 101129732 and DOI 10.3030/101129732.

The authors would like to thank all the institutions and their representatives for their collaboration in the exchange of data to the EHRI Portal.
\end{acks}

\bibliographystyle{ACM-Reference-Format}
\bibliography{references}

\end{document}